# DISCUSSION OF: TREELETS—AN ADAPTIVE MULTI-SCALE BASIS FOR SPARSE UNORDERED DATA

By Xing Qiu

*University of Rochester*

This is a discussion of paper "Treelets—An adaptive multi-scale basis for sparse unordered data" by Ann B. Lee, Boaz Nadler and Larry Wasserman. In this paper the authors defined a new type of dimension reduction algorithm, namely, the treelet algorithm. The treelet method has the merit of being completely data driven, and its decomposition is easier to interpret as compared to PCR. It is suitable in some certain situations, but it also has its own limitations. I will discuss both the strength and the weakness of this method when applied to microarray data analysis.

**1. The design of the treelet algorithm.** A lot of modern technologies require analyzing noisy, high-dimensional and unordered data. As an example, in the field of microarray analysis, researchers are often interested in analyzing gene expessions sampled from $n$ different subjects. These expression data can be seen as $n$ independent realizatons of a $p$-dimensional random vector $\vec{x} = (x_1, \ldots, x_p)^T$, each $x_i$ represents (usually log tranformed) an expression level of a given gene. In practice, $p$ (number of genes) is measured in thousands or tens of thousands, and $n$ (sample size) is more than often less than a dozen. Due to this "large $p$, small $n$" nature, dimension reduction such as hierarchical clustering (denoted as HC henthforth) is often conducted prior to regression or classification analysis.

The treelet algorithm can be best described as a data driven local PCA (Principal Component Analysis). It can be summarized in the following steps:

1. Find the two most similar variables (genes) by a well-defined metric of similarity such as covariance. Denote this pair of genes as $x_\alpha$, $x_\beta$.
2. Perform a local PCA on this pair to decorrelate them. More specifically, find a Jacobi rotation matrix $J$ such that $x^{(2)} = J^T(x)$ has this property:

$\operatorname{cov}(x_\alpha^{(2)}, x_\beta^{(2)}) = 0$. Then drop the less important one of them (the one with smaller variance) and update the similar matrix.

In other words, after this step, a summary variable will be chosen to replace the two most similar variables from the original data.
3. Update the similarity matrix with this new summary variable and then find the next most similar pair of variables.
4. Build up a multi-resolution analysis accordingly. At each step, we have a representation of $x$ as the sum of the coarse-grained representation of the signal and the sum of the residuals.

**2. Comparisons to other methods.** Dimension reduction is not a new technique in data analysis. Principal component analysis [Jolliffe (2002)] and hierarchical clustering methods [Eisen et al. (1998), Tibshirani et al. (1999)] are among the most used methods in this arena.

PCA As pointed out by the authors, PCA computes a *global* representation of data. The principal components are linear combinations of all variables. This poses an obstacle for interpreting the results. On the other hand, the treelet method is a *local* method by design. For example, when the underlying dependence structure of data can be modeled as disjoint groups of variables which are uncorrelated to each other groupwise, in principle, local dimension reduction methods should perform better than their global counterpart.

HC In a sense, the treelet can be viewed as yet another way of constructing the dendrogram from the bottom up. So the treelet method is a legitimate member of the family of agglomerative hierarchical clustering algorithms. However, there is a novelty in the treelet method approach. By construction, at each step only the sum variable (the variable which contributes more variance) remains as the representative of the pair of closely related variables. At the end of the day, the dendrogram will reflect the "skeleton" of the given data rather than the dependence structure of the data themselves. If in a specific application we have evidence that the unused residual terms reflect nothing but noise, then the treelet method provides us invaluable information about hierarchical dependence of the data which is noise resistant.

**3. Applicability in the field of microarray data analysis.** As mentioned in Section 1, microarray data analysis is a good example where the treelet method may shine. It is a well-known biological fact that genes work together instead of independently. As a consequence, their expressions are highly correlated.

Storey and Tibshirani (2003) hypothesized that most likely the form of intergene dependence is *weak dependence*, which can be "...loosely described as any form of dependence whose effect becomes negligible as the number of features increases to infinity." And their argument is that genes can be grouped into essentially independent *pathways*.

If this hypothesis is true, then the treelet method would work beautifully, as illustrated in Chapter 3.2 of Lee, Nadler and Wasserman (2008).

However, a series of study conducted by Qiu et al. (2005a, 2005b, 2006) on St. Jude Children's Research Hospital Database (see sjcrh database on childhood leukemia) showed that, on average, the intergene correlation level is too high to be explained by the within pathway dependence (weak dependence) alone. There is strong long ranged *global* dependence between pathways. Whether this global dependence is due to technical noise or not is up to debate [Klebanov and Yakovlev (2007)]. If the observed high intergene correlation is due to biological reasons rather than noise, then the treelet method may be harmful since it will reduce and distort useful information contained in the dependence structure.

It is also interesting to compare the treelet method with various normalization methods, such as the global normalization [Yang et al. (2002)]. Apparently, global normalization (or any other normalization method) is not a dimension reduction procedure, nor does it give us a dendrogram. However, one similarity can be found between the global normalization and the treelet method: they both replace data variables with surrogate variables which are linear combinations of the original variables. In the case of global normalization (assuming expression levels are log transformed), the $i$th variable (gene) $x_i$ is replaced by $x_i - \bar{x}$, where $\bar{x}$ is the sample average of $x$ over all genes for a given slide. From this point of view, global normalization is a *global* basis transformation. A hidden assumption in doing global normalization is that $\bar{x}$ represents slide-specific noise thus needs to be removed from the observed signal. While I personally think that technical noise cannot be removed in such an overly simplistic way, it provides an example where a global method may better capture the most useful information at a much faster rate.

Another dangerous behavior of the treelet method is that it uses variance as a means to evaluate which variable should be retained (sum variable), and which one should be disregarded (difference variable). This approach may look very plausible mathematically, yet it ignores the possibility that genes with lesser variability may actually be the important ones. It may very well be the case that in evolution genes that are responsible for essential functionalities are more likely to have smaller variation than those less important ones.

One of the major advantage of the treelet method is that the sum variables it produces use only a subset of variables. This makes it easier to interpret than PCR, which gives linear combinations of *all* variables as outcome. However, the sum variables of the treelet method can also be linear combinations of *many* variables. It is a huge leap forward in the right direction, yet it is still hard to find its way into another important field of microarray analysis: testing differential expressions. Being hard to interpret is just an apparent disadvantage. A more subtle disadvantage is that there is no guarantee that the multiple testing procedures designed to work with original expressions still control the same false positive level when we replace them with some "noise-free" surrogate variables. Much future work can be done in this direction.

**4. Discussion.** Overall, I think the treelet method has the merit of being completely data driven and being local. I am very impressed by its performance when data variables are divided into uncorrelated groups.

However, when talking about its applicability to gene expression data, I think a lot of careful investigation still needs to be done. This is due to the complexity of the dependence structure exhibits in this type of data. This complexity is probably the reason why the treelet method (in its original form) did not outperform other classification methods on the leukemia data set of Golub et al.

In the future more attention should be paid to the nature of inter-pathway dependence. Should we model pathways as disjoint, uncorrelated "super variables"? Or should we also model some long range, inter-pathway correlation? I think this question can be answered only through joint efforts from both statisticians and biologists.

I also want to point out that I disagree with the authors in that PCA cannot reveal the underlying noiseless structure of the data while the treelet method can. As pointed out by numerous researchers [Storey et al. (2007), Barbujani et al. (1997), Akey et al. (2002), Rosenberg et al. (2002)], most human genetic variation is due to variation among individuals within a population rather than among populations. This implies that the majority of "noise" in the data is actually true biological information. So being too good at removing "noise" may not always be a merit.

In short, I believe there is no one-size-fits-all solution for noisy, high-dimensional data. The treelet method provides us a very good solution in some situations, and it opens many research possibilities in the future.

Possible future improvements:

- The leukemia data set of Golub et al. used for classification of DNA microarray data is not the largest data available. The authors may want to try St. Jude Children's Research Hospital Database on childhood leukemia too.

- In the same chapter, the authors claim that they use a novel "two-way treelet decomposition scheme." They first compute treelets on the genes, then compute treelets on the *samples*. It looks very suspicious. I have a feeling that the gained performance is due to some subtle violation of the principle of external cross-validation. The authors should definitely provide more details about this approach.
- A recent paper by Klebanov, Jordan and Yakovlev (2006) proposed a new model of the long range intergene correlation structure. In a loose way, they hypothesize that there exist "gene drivers" and "gene modulators," such that the expression of a "gene-modulator" is stochastically proportional to that of a "gene-driver" (without log transformation). It would be nice to see if the treelet method works in this situation.

## REFERENCES

St. Jude children's research hospital (sjcrh) database on childhood leukemia.
Akey, J. M., Zhang, G., Zhang, K., Jin, L. and Shriver, M. D. (2002). Interrogating a high-density snp map for signatures of natural selection. *Genome Res.* **12** 1805–1814.
Barbujani, G., Magagni, A., Minch, E. and Cavalli-Sforza, L. L. (1997). An apportionment of human dna diversity. *Proc. Natl. Acad. Sci. USA* **94** 4516–4519.
Eisen, M., Spellman, P., Brown, P. and Botstein, D. (1998). Cluster analysis and display of genome-wide expression patterns. *Proc. Natl. Acad. Sci. USA* **95** 14863–14868.
Jolliffe, I. T. (2002). *Principal Component Analysis*. Springer, New York. MR2036084
Klebanov, L., Jordan, C. and Yakovlev, A. (2006). A new type of stochastic dependence revealed in gene expression data. *Stat. Appl. Genet. Mol. Biol.* **5** Article 7. MR2221298
Klebanov, L. and Yakovlev, A. (2007). How high is the level of technical noise in microarray data? *Biol. Direct.* **2** 9.
Lee, A. B., Nadler, B. and Wasserman, L. (2008). Treelets—An adaptive multi-scale basis for sparse unordered data. *Ann. Appl. Statist.* To appear.
Qiu, X., Brooks, A. I., Klebanov, L. and Yakovlev, A. (2005a). The effects of normalization on the correlation structure of microarray data. *BMC Bioinformatics* **6** 120.
Qiu, X., Klebanov, L. and Yakovlev, A. (2005b). Correlation between gene expression levels and limitations of the empirical bayes methodology in microarray data. *Statist. Appl. Genet. Mol. Biol.* **4** Article 3. MR2183944
Qiu, X. and Yakovlev, A. (2006). Some comments on instability of false discovery rate estimation. *J. Bioinformatics Computational Biology* **4** 1057–1068.
Rosenberg, N. A., Pritchard, J. K., Weber, J. L., Cann, H. M., Kidd, K. K., Zhivotovsky, L. A. and Feldman, M. W. (2002). Genetic structure of human populations. *Science* **298** 2381–2385.
Storey, J. D. and Tibshirani, R. (2003). Statistical significance for genomewide studies. *Proc. Nat. Acad. Sci. USA* **100** 9440–9445. MR1994856
Storey, J. D., Madeoy, J., Strout, J. L., Wurfel, M., Ronald, J. and Akey, J. M. (2007). Gene-expression variation within and among human populations. *Am. J. Hum. Genet.* **80** 502–509.
Tibshirani, R., Hastie, T., Eisen, M., Ross, D., Botstein, D. and Brown, P. (1999). Clustering methods for the analysis of dna microarray data. Technical report, Dept. Statistics, Stanford Univ.

Yang, Y. H., Dudoit, S., Luu, P., Lin, D. M., Peng, V., Ngai, J. and Speed, T. P. (2002). Normalization for cdna microarray data: A robust composite method addressing single and multiple slide systematic variation. *Nucleic Acids Res.* **30** e15.

Department of Biostatics
  and Computational Biology
University of Rochester
601 Elmwood Ave
BOX 630
Rochester, New York 14642
USA
E-mail: xqiu@bst.rochester.edu